**Plasmons in ballistic nanostructures with stubs: transmission line approach**

G. R. Aizin[1], J. Mikalopas[1], and M. Shur[2]

[1] Kingsborough College. The City University of New York, Brooklyn, NY, 11235, USA

[2] Rensselaer Polytechnic Institute, Troy, NY 12180, USA

ABSTRACT

The plasma wave instabilities in ballistic Field Effect Transistors (FETs) have a promise of developing sensitive THz detectors and efficient THz sources. One of the difficulties in achieving efficient resonant plasmonic detection and generation is assuring proper boundary conditions at the contacts and at the heterointerfaces and tuning the plasma velocity. We propose using the tunable narrow channel regions of an increased width, which we call "stubs" for optimizing the boundary conditions and for controlling the plasma velocity. We developed a compact model for THz plasmonic devices using the transmission line (TL) analogy. The mathematics of the problem is similar to the mathematics of a TL with a stub. We applied this model to demonstrate that the stubs could effectively control the boundary conditions and/or the conditions at interfaces. We derived and solved the dispersion equation for the device with the stubs and showed that periodic or aperiodic systems of stubs allow for slowing down the plasma waves in a controllable manner in a wide range. Our results show that the stub designs provide a way to achieve the optimum boundary conditions and could also be used for multi finger structures – stub plasmonic crystals - yielding better performance of THz electronic detectors, modulators, mixers, frequency multipliers and sources.

I. INTRODUCTION

Numerous existing and potential applications of THz technology require efficient electronic THz sources and detectors. [1] This is especially important for THz communications supporting beyond 5G WI FI that would require massive deployment of sub-THz and THz systems. Developing the THz electronic sources is the key challenge to be met for closing the famous THz gap. The existing sub-THz and THz electronic sources use Gunn [2] and IMPATT [3] diodes (directly or with frequency multiplication by Schottky diodes [4]), InP based High Electron Mobility Transistor [5] or Si CMOS [6] and BiCMOS [7] Integrated Circuits. The plasma wave instabilities in ballistic Field Effect Transistors (FETs) have a promise of enabling more efficient and tunable THz sources based on the Dyakonov-Shur (DS) instability [8] and the "plasmonic boom" instabilities [9,10].

The mechanism of the DS instability involves the plasma wave reflections from the source and drain channel edges. The DS instability has the largest increment when the boundary conditions at the source and drain edges of the device channel correspond to zero amplitude and to the largest amplitude of the THz electric field variation, respectively. It has not been clear how to realize such boundary conditions at the THz frequencies.

The problem of controlling the boundary conditions and conditions at interfaces between different device sections becomes even more important for the periodic "plasmonic crystal" structures using the sections experiencing the DS instability [11] or "plasmonic boom" transitions. [9, 10] In the latter structures, the plasmonic instability occurs when the electron drift velocity crosses the plasma wave velocity. This could be achieved by modulating the carrier concentration in the plasmonic device channel and/or the channel width. In either case, controlling the conditions at the interfaces of the channel sections are very important.

The "plasmonic boom" mechanism requires the electron drift velocity to become higher than the plasma velocity. This requirement might be difficult to meet because the plasma velocity is typically quite high. As shown in this paper, using stubs would allow to slow down the plasma waves in a controllable fashion.

The plasmonic detection/generation experiments reported so far relied on the asymmetry at the source and drain contact. Such asymmetry is sufficient, in principle, to achieve the THz detection or even the DS instability but the sensitivity of the detection or the instability increment are enhanced in the structure with the build-in asymmetry [12].

In this paper, we describe how to solve these problems by adding narrow protruding regions with tunable carrier concentration to the device channel (see Fig. 1). Following [13], we call these regions "stubs." This term emphasizes the analogy with the transmission line formalism that has been used for the analysis of the plasmonic structures. [14, 15, 16, 17, 18]

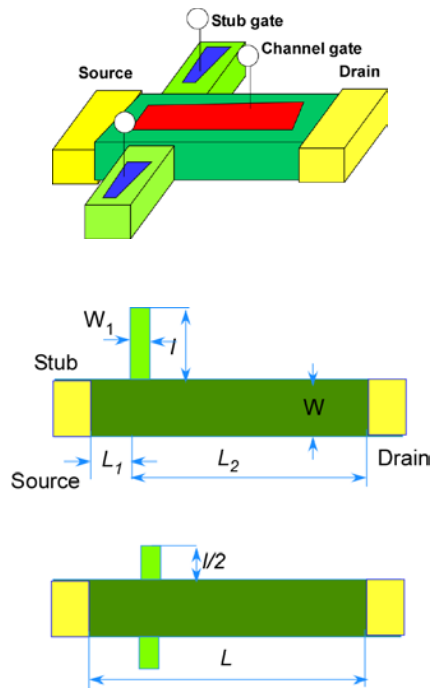

Figure 1. Stub configurations and notations.

Fig. 2 shows the designs with different implementations of tuning the effective electric properties of the stub. The top gate controls the carrier concentration in the stub. The side gate implementation [19] allows for an effective control of the stub width and length. The combination of the top and side gates could also suppress the narrow and/or short channel effects.

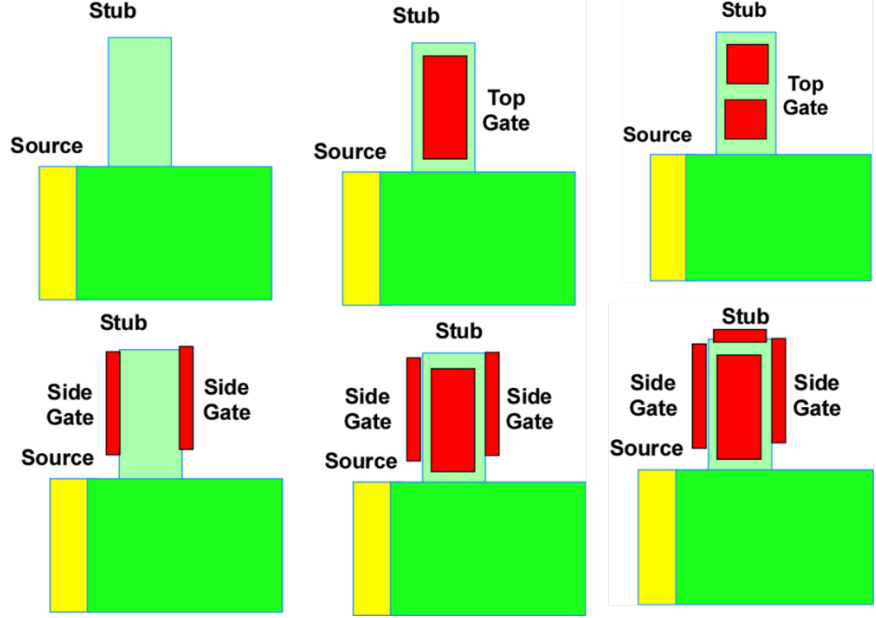

Figure 2. Unmodulated stub and different stub modulation schemes using top and side gates.

The stub introduces impedance $Z_{st}$ that could vary from minus infinity to infinity at the plasma frequency depending on the stub effective area and other parameters. Hence, placing the stubs at the two edges of the channel should allow for setting the optimum boundary conditions for the DS instability. Introducing the stubs at the region interfaces should allow for controlling the transitions between the interfaces and phenomena such as choking [20].

Our results also show that stubs could slow down the plasma waves, which is important for achieving the plasmonic boom conditions [9, 10].

2. BASIC EQUATIONS

We consider plasma waves propagating between the source and the drain contacts (*x*-axis) in the gated 2D electron layer located at the $z = 0$ plane.

The hydrodynamic equations (the Euler equation and equation of continuity) for the local sheet electron density $n(x, t)$ and velocity $v(x, t)$ in the plasma wave are

$$\frac{\partial v}{\partial t} + v \frac{\partial v}{\partial x} = \frac{e}{m^*} \frac{\partial \varphi}{\partial x}, \tag{1}$$

$$\frac{\partial n}{\partial t} + \frac{\partial (nv)}{\partial x} = 0, \tag{2}$$

where $\varphi(x, z = 0, t)$ is electric potential in the 2D layer, $-e$ and $m^*$ are the electron charge and effective mass, respectively. The hydrodynamic approach is justified if the time of the electron-electron collisions is much less than the time of electron collisions with impurities and phonons and the electron transit time across the device. In the Euler equation, we omitted the pressure gradient term since in the gated 2D channels this term is typically much smaller than the field term in the above threshold regime as was discussed in [10].

We neglect effect of collisions with phonons and impurities on the electron transport, i.e. we assume the ballistic electron transport with the electron mean free path larger than the channel length and $\omega\tau \gg 1$, $\omega$ is the plasmon frequency, $\tau$ is characteristic collision time.

Equations (1, 2) could be linearized with respect to the small fluctuations of the electron density $\delta n(x,t)$ and velocity $\delta v(x,t)$ by assuming that $n(x,t) = n_s + \delta n$ where $n_s$ is the equilibrium electron density [21]

$$n_s = n_0 \ln\left(1 + e^{\frac{eU_{gt}}{\eta k_B T}}\right) \tag{3}$$

Here $U_{gt}$ is the gate voltage swing, $k_B$ is the Boltzmann constant, $\eta$ is the ideality factor, $T$ is temperature, $n_0 = \frac{C_0 \eta k_B T}{e^2}$, $C_0 = \frac{\varepsilon_0 \varepsilon}{d + \Delta d}$ is the gate-to-channel capacitance per unit area, $\varepsilon$ is the dielectric constant, $\varepsilon_0$ is the dielectric permittivity of vacuum, $d$ is the channel-to-gate separation, and $\Delta d$ is the effective thickness of the 2DEG.

We consider the strong gate screening limit when $\delta n$ is linked with the fluctuation of electric potential $\delta\varphi$ as $-e\delta n = C\delta\varphi$, where $C$ is the differential capacitance given by

$$C = \frac{C_0}{1 + e^{-\frac{eU_{gt}}{\eta k_B T}}} \tag{4}$$

In this case, the solution of the linearized equations for the Fourier harmonics ($\delta n, \delta v \propto e^{-iqx+i\omega t}$) is

$$I_\omega(x) = I_1 e^{-iq_1 x} + I_2 e^{-iq_2 x} \tag{5}$$

$$V_\omega(x) = \frac{1}{CWv_p}\left(I_1 e^{-iq_1 x} - I_2 e^{-iq_2 x}\right) \tag{6}$$

where $I_\omega = -eWn_s \delta v$ is the total current and $V_\omega \equiv \delta\varphi_\omega$ is the voltage distribution in the plasma wave of frequency $\omega$ propagating in the channel of width $W$, $q_{1,2} = \pm \omega/v_p$ are the wave vectors of the plasma waves propagating in two opposite directions in the channel, and

$$v_p = \sqrt{\frac{\eta k_B T}{m^*}\left(1 + e^{-\frac{eU_{gt}}{\eta k_B T}}\right)\ln\left(1 + e^{\frac{eU_{gt}}{\eta k_B T}}\right)} \tag{7}$$

is the plasma velocity. [22]. Constants $I_1$ and $I_2$ are determined by the boundary conditions.

As discussed in [14-18], description of the plasma waves in the 2D electron system within the hydrodynamic model is analogous to the description of the electromagnetic signals in the transmission line (TL). The linearized hydrodynamic equations for the plasma waves in the 2D electron channel are equivalent to the telegrapher's equations for the TL with the distributed inductance $\mathcal{L} = \frac{m^*}{e^2 n_0 W}$, resistance $\mathcal{R} = \mathcal{L}/\tau$ per unit channel length, and distributed capacitance, which depends on the gating conditions [14, 17]. In the limit of a strong gate screening, the distributed capacitance per unit length equals to $CW$ with $C$ defined in (4). The 2D electron channel can be viewed as a plasmonic waveguide supporting transverse electromagnetic (TEM) plasma modes. FET with the stubs with separately biased gates allows for tuning the stub impedances by adjusting the carrier concentration in the stubs and, therefore, the plasma velocity in the FET channel (see Fig. 2).

Within the TL approach, the stub in Fig. 1 represents an open circuit TL stub maintaining its own standing plasma modes and characterized by the input stub impedance

$$Z_{st} = -iZ_{01} \cot \frac{\omega l}{v_{p1}} \qquad (8)$$

where $Z_{01} = 1/(C_1 W_1 v_{p1})$ is the characteristic impedance of the plasmonic TL in the stub, $l$ and $W_1$ are the length and the width of the stub, $C_1$ and $v_{p1}$ are the differential capacitance and plasma velocity in the stub defined by Eq. (4) and (7), respectively. This stub model is valid if the width of the stub $W_1$ is much smaller than the plasmon wavelength in the 2D FET channel so that the junction of the stub to the channel is a well-defined point. Zeros of $Z_{st}$ correspond to the excitations of the standing plasma modes in the stub with wavelengths $\lambda_n = \frac{4l}{(2n+1)}$, $n = 0,1,\ldots$ [13].

Fig. 3 shows the equivalent TL electric circuit for the FET shown in Fig. 1 with the stub located close to the source edge of the channel (Fig. 3a) and at some arbitrary position in the channel at distances $L_1$ and $L_2$ from the source and the drain (Fig. 3b).

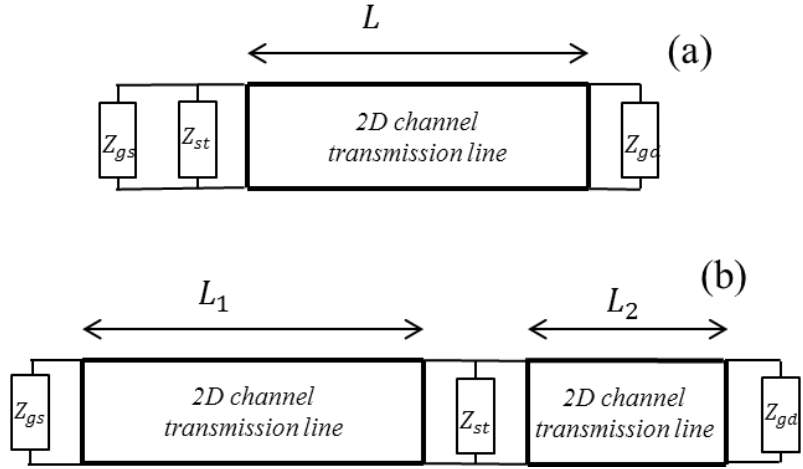

Figure 3. Equivalent TL electric circuits of the FET channel with a stub.

In this Figure, we also introduced terminating impedances $Z_{gs}$ and $Z_{gd}$ describing the electric links between the gate and the source and drain contacts, respectively. These links determine the boundary conditions in the plasmonic cavity formed between the source ($x = 0$) and the drain ($x = L$) contacts in the FET as

$$V_\omega(0) = -Z_{gs}I_\omega(0) \tag{9}$$
$$V_\omega(L) = Z_{gd}I_\omega(L)$$

It follows from (8) that the value of $Z_{st}$ at the plasma frequency can be tuned from minus infinity to infinity by varying the carrier concentration and the geometry of the stub. If the stubs are located near the source and/or the drain edge of the channel as shown in Fig. 3a the impedance $Z_{st}$ is combined with $Z_{gs}$ and $Z_{gd}$ to provide a very effective way to control and optimize the boundary conditions for the plasmonic FET.

To illustrate this conclusion, we used the TL formalism to calculate the input gate-to-source impedance $Z_{in}$ in the stub FET assuming $Z_{gd} = \infty$:

$$Z_{in} = -iZ_0 \frac{\cos\frac{\omega(L_1+L_2)}{v_p} - \frac{Z_0}{Z_{01}}\tan\frac{\omega l}{v_{p1}}\sin\frac{\omega L_1}{v_p}\cos\frac{\omega L_2}{v_p}}{\sin\frac{\omega(L_1+L_2)}{v_p} - \frac{Z_0}{Z_{01}}\tan\frac{\omega l}{v_{p1}}\cos\frac{\omega L_1}{v_p}\cos\frac{\omega L_2}{v_p}}. \tag{10}$$

Here $Z_0 = 1/WCv_p$ is the characteristic impedance of the plasmonic TL in the FET 2D channel.

Fig. 4 shows the calculated dependence of $Z_{in}/iZ_0$ on the frequency $\omega/\omega_0$, $\omega_0 = \frac{v_p}{L}$, for the stub FET with a stub located close to the source ($L_1 \ll L_2$) and $Z_{01} = Z_0$:

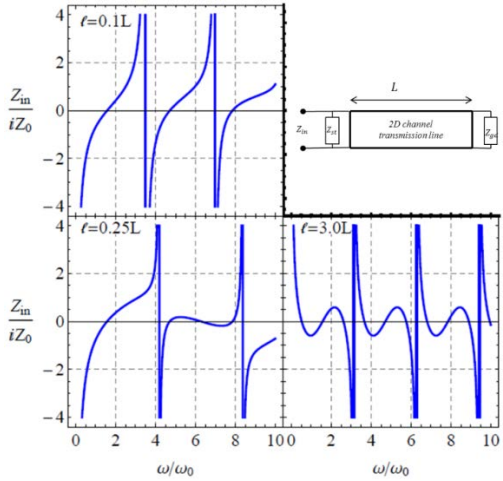

Figure 4. Frequency dependence of the input source-to-gate impedance for the stub FET shown in Fig.1 at different stub lengths $l$. Inset: equivalent TL electric circuit.

As seen from (8) and Fig. 4, the input impedance is tunable in a wide range by changing the parameters of the stub, such as the electron density or effective dimensions of the stub using the stub modulation schemes shown in Fig. 2. The impedance derivative with respect to frequency could be also tuned as clearly seen from Fig. 4, where this derivative is zero at certain frequencies. This result is important for designing broad band plasmonic devices.

In agreement with the results reported in [13], Fig. 4 shows that the system with a stub has two types of plasmonic resonances ($Z_{in} = 0$): (1) the resonances associated with the channel and (2) the resonances associated with the stub. The latter resonances are tunable by varying the stub parameters. Anti-resonances ($Z_{in} = \infty$) occur when the stub and the channel impedances (that are connected in parallel, see the inset in Fig. 4) cancel each other. Tuning the plasmonic frequencies and the input impedance in a wide range by varying the effective stub parameters enables the applications of the system with stubs for the terahertz spectroscopy by adjusting the resonance frequency to coincide with the frequency of the impinging THz signal. Another application is in the resonant THz interferometry. [23]

Whereas the stubs located near the source and drain ends of the channel allow us to control the input and output impedances and, hence, the boundary conditions, the stubs attached to the channel at some distance from the source and the drain edges as shown in Fig. 3b control the plasmonic spectrum of the channel.

To demonstrate this effect, we derived the plasmon dispersion relation for the plasmonic cavity with the stub shown in Fig. 3b. We used the TL formalism to solve (5) with the stub impedance given by (8) and the boundary conditions given by (9) with the result

$$e^{i\frac{2\omega L_2}{v_p}} = \frac{\left(1 - \frac{Z_0}{Z_{gd}}\right)\left(1 - \frac{Z_0}{Z_{st}} - \frac{Z_0}{Z}\right)}{\left(1 + \frac{Z_0}{Z_{gd}}\right)\left(1 + \frac{Z_0}{Z_{st}} + \frac{Z_0}{Z}\right)}$$

(11)

$$Z = Z_0 \frac{Z_{gs}\cos\frac{\omega L_1}{v_p} + iZ_0\sin\frac{\omega L_1}{v_p}}{Z_0\cos\frac{\omega L_1}{v_p} + iZ_{gs}\sin\frac{\omega L_1}{v_p}},$$

In Fig. 5, we present the results of the numerical solution of the plasmon dispersion equation (11) in case when the stub is positioned at the center of the channel, $L_1 = L_2 = L/2$, and $v_p = v_{p1}$, $W_1 = W$. Plasmonic spectrum is shown as a function of the stub length $l$ assuming $Z_{gs} = 0$ and $Z_{gd} = \infty$. As seen from Fig. 5, addition of the stub reduces the unperturbed resonant plasmonic frequencies $\omega_n = \frac{\pi v_p}{2L}(2n+1)$, $n = 0,1,...$ for the same channel length and, hence, for the same wave vector. This means the plasma waves in the device with a stub have a smaller velocity. The unique feature of the stub design is that this velocity is tunable by the stub side and/or top gates modulating the effective stub area and/or the carrier concentration in the stub (see Fig. 2). For the conventional design, controlling the carrier concentration in the channel simultaneously adjusts the plasma velocity and the input impedance. The addition of the independent stub gate control allows for an optimization of the plasmonic structures for THz electronics applications independently adjusting the plasma velocity and the input and output impedances.

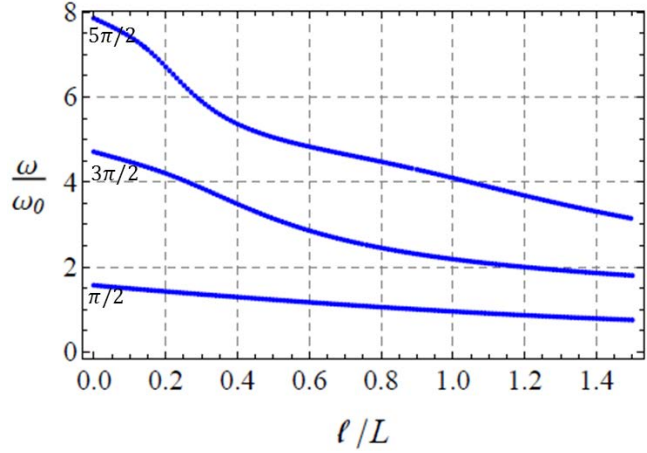

Figure 5. Plasmon frequencies in the stub FET plasmonic cavity of length $L$ with a stub at the center of the cavity as a function of the stub length $l$.

Inserting the stubs near the source and/or near the drain separates the channel design from the input/output impedance design. The source stub could control $Z_{in}$, The drain stub could control $Z_{out}$. One or more stubs attached to the channel could control the channel properties. This approach makes the stubs a perfect optimization tool box for the plasmonic design.

The generalization of this approach are stub plasmonic crystals of different dimensionality and stub plasmonic crystals with complex (asymmetric) elementary cell shown in Figs. 6a, 6b, 6c and 6d, respectively. Fig. 6e shows the equivalent TL electric circuit for a 1D stub plasmonic crystal in Fig. 6a.

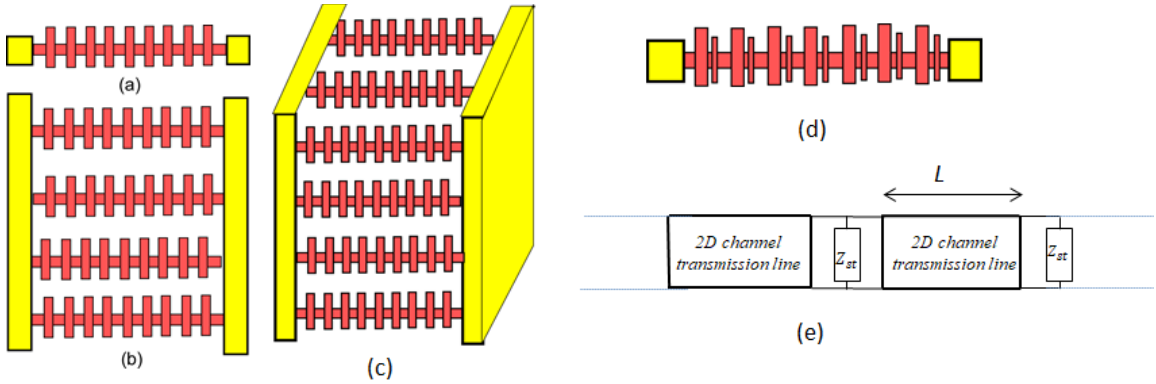

Figure 6. Schematic diagram of the stub plasmonic crystals: (a) 1D; (b) 2D, (c) 3D (d) 1D asymmetric stub plasmonic crystal; (e) equivalent TL electric circuit for the 1D stub plasmonic crystal

The "Kronig-Penney" analysis yields the following plasmon dispersion equation for the 1D stub plasmonic crystal with period $L$:

$$\cos kL = \cos\frac{\omega L}{v_p} - \frac{Z_0}{2Z_{01}}\tan\frac{\omega l}{v_{p1}}\sin\frac{\omega L}{v_p} \tag{12}$$

where $k$ is the plasmon wave vector. In the first fundamental band at $k \to 0$ the plasmon dispersion law is linear with plasmon velocity $v_{pl}$ given by

$$v_{pl} = \pm \frac{v_p}{\sqrt{1+\frac{Z_0 l}{Z_{01} L}}} \qquad (13)$$

It follows from (13) that plasmons are slowing down, and this process can be tuned by adjusting the effective width or length, or effective carrier concentration in the stubs. Fig. 7 shows the plasmon dispersion relations calculated from (12) for stubs of different length and width and $v_p = v_{p1}$ illustrating this conclusion.

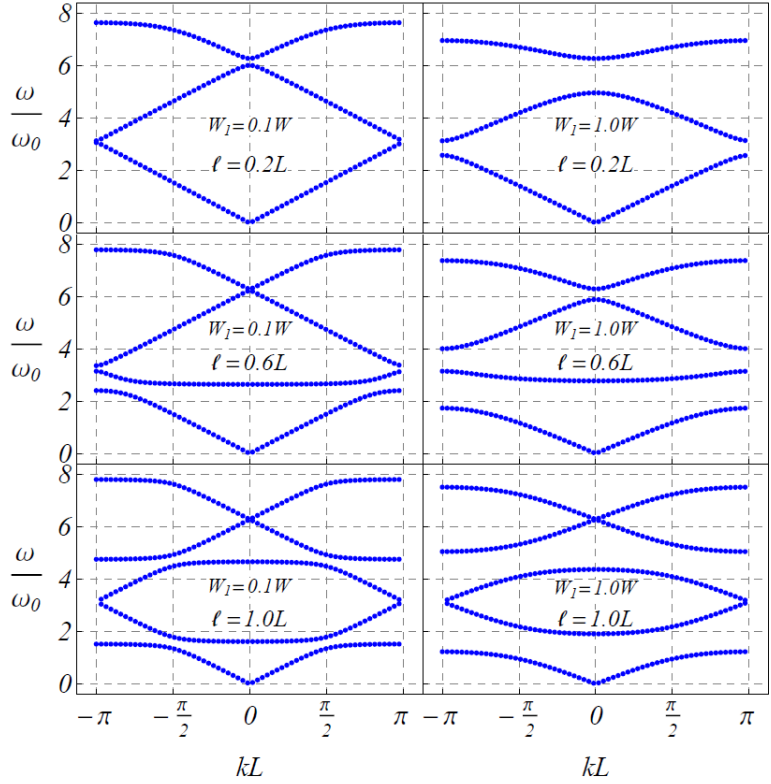

Figure 7. Plasmonic band spectrum in the 1D stub plasmonic crystal with period $L$ at different lengths $l$ and widths $W_1$ of the stub. $W$ is the width of the 2D channel.

In this case, (13) reduces to

$$v_{pl} = \pm \frac{v_p}{\sqrt{1+s}} \qquad (14)$$

where $s = W_1 l / W L$ is the effective stub area equal to the ratio of the stub area to the area of the one period of the 2D channel. Fig. 8 shows variation of the plasma velocity $v_{pl}$ with the effective stub area $s$ and the derivative $\left|\frac{dv_{pl}}{ds}\right|$ that shows a high sensitivity of the plasmon velocity to the effective stub area that could be adjusted by the stub gate bias, see Fig. 2. This is very important for the optimization of the plasmonic THz devices.

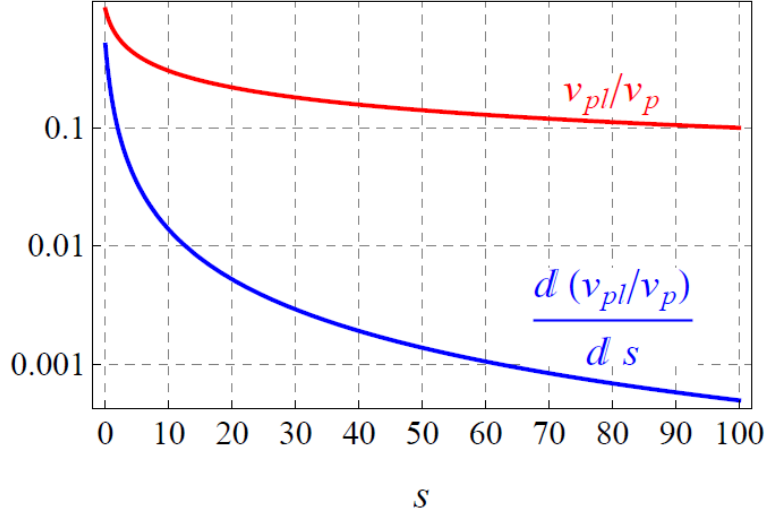

Figure 8. Plasmon velocity $v_{pl}$ (red) and $|dv_{pl}/ds|$ (blue) as a function of the effective stub area, $s$, in the 1D plasmonic stub crystal

Varying the stub area in the sequence of stubs (as shown in Fig. 9) allows for a gradual change of the plasma velocity along the channel.

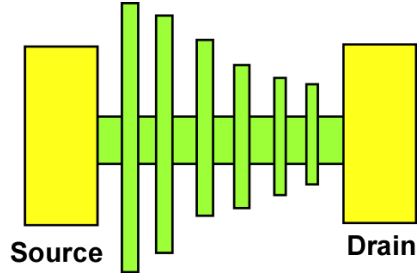

Fig. 9. Sequence of stubs enabling a gradual change of the plasma velocity along the channel

The above results for the plasmonic crystal had been derived assuming ballistic electron transport when scattering of electrons on phonons and impurities can be neglected. Accounting for the finite momentum relaxation time, $\tau$, modifies the dispersion equation (12) as follows

$$\cos kL = \cos\frac{\omega\sqrt{1-\frac{i}{\omega\tau}}L}{v_p} - \frac{W_1}{2W}\tan\frac{\omega\sqrt{1-\frac{i}{\omega\tau}}l}{v_p}\sin\frac{\omega\sqrt{1-\frac{i}{\omega\tau}}L}{v_p} \qquad (15)$$

For the most interesting case when $\omega\tau \gg 1$, (15) becomes

$$\cos kL = \cos\left(\frac{\omega L}{v_p} - \frac{iL}{2v_p\tau}\right) - \frac{W_1}{2W}\tan\left(\frac{\omega L}{v_p} - \frac{iL}{2v_p\tau}\right)\sin\left(\frac{\omega L}{v_p} - \frac{iL}{2v_p\tau}\right) \qquad (16)$$

As seen, the key parameter for the ballistic transport approximation is $\frac{v_p\tau}{L}$. This parameter has to exceed unity for the quasi ballistic approximation to be valid. For typical room temperature values of $\tau \sim 10^{-13}$ s and $v_p \sim 10^6$ m/s, we obtain $L < 100$ nm. This is a very realistic value that could be exceeded by one or even two orders of magnitude at cryogenic temperatures and in advanced materials, such as graphene.

The proposed approach enables a large variety of the tunable THz devices ranging from sources, to spectrometers, interferometers, mixers, frequency multipliers, modulators, and detectors. In the latter case, the design with asymmetric stubs (as schematically shown in Fig. 6d is beneficial.

3. CONCLUSIONS

The device impedance in the presence of stubs has a wide range tunability that can be used for optimizing the plasmonic detectors and sources. The solutions of the dispersion equation for the device with the stubs demonstrated that the stubs can adjust and tune the plasma velocity. This feature makes the plasmonic boom devices feasible.

This new device design also allows to implement 1D, 2D, and 2D plasmonic crystals for achieving higher powers and better radiation extraction.

The proposed TL modeling approach support compact design modeling of the future generation of plasmonic THz devices for applications in THz sensing, imaging, and communications.

4. ACKNOWLEDGMENTS

The work at RPI and CUNY was supported by the US Army Research Office (Project Manager Dr. Joe Qiu).